\documentclass[a4paper,10pt]{article}
\usepackage{graphicx}
\usepackage{comment}
\usepackage{amssymb}
\usepackage{amsmath}
\usepackage{nicefrac}
\usepackage{graphicx}
\usepackage{dcolumn}
\usepackage{bm}
\usepackage{comment}

\begin{document}

\newcommand{\bin}[2]{\left(\begin{array}{c} \!\!#1\!\! \\  \!\!#2\!\! \end{array}\right)}
\newcommand{\troisj}[3]{\left(\begin{array}{ccc}#1 & #2 & #3 \\ 0 & 0 & 0 \end{array}\right)}
\newcommand{\troisjm}[6]{\left(\begin{array}{ccc}#1 & #2 & #3 \\ #4 & #5 & #6 \end{array}\right)}
\newcommand{\sixj}[6]{\left\{\begin{array}{ccc}#1 & #2 & #3 \\ #4 & #5 & #6 \end{array}\right\}}
\newcommand{\neufj}[9]{\left\{\begin{array}{ccc}#1 & #2 & #3 \\ #4 & #5 & #6 \\ #7 & #8 & #9 \end{array}\right\}}
\newcommand{\eline}{E_{\gamma J,\gamma'J'}}

\huge

\begin{center}
Description of anomalous Zeeman patterns in stellar astrophysics
\end{center}

\vspace{0.5cm}

\large

\begin{center}
Jean-Christophe Pain\footnote{jean-christophe.pain@cea.fr} and Franck Gilleron
\end{center}

\normalsize

\begin{center}
\it CEA, DAM, DIF, F-91297 Arpajon, France
\end{center}

\vspace{0.5cm}

\begin{abstract}
The influence of a magnetic field on the broadening of spectral lines and transition arrays in complex spectra is investigated. The anomalous absorption or emission Zeeman pattern is a superposition of many profiles with different relative strengths, shifts, widths, asymmetries and sharpnesses. The $\sigma$ and $\pi$ profiles can be described statistically, using the moments of the Zeeman components. We present two statistical modellings: the first one provides a diagnostic of the magnetic field and the second one can be used to include the effect of a magnetic field on simulated atomic spectra in an approximate way. 
\end{abstract}

\section{Introduction}
The existence of intense magnetic fields (1 - 1000 MG) for white dwarfs was confirmed by the observation of the splitting of spectral lines in the visible and UV range. Pulsars and neutron stars, discovered through their spectrum in the range of radio-frequencies and X-rays, have an even more intense magnetic field. Magnetic fields of a few kG also affect the radiative accelerations in chemically peculiar (CP) stars [Alecian \& Stift (2002)]. The emission or absorption lines of an atom are modified by the magnetic fields. Thanks to the Zeeman effect [Zeeman (1897)], such fields can be detected at large distances, through the measured radiation. The methods differ according to the nature of the stars studied, the magnitude and the geometry of the magnetic fields and the spectral resolution. Moreover, the variations of the magnetic field of stars during their rotation bring some information about their global geometry. In spectro-polarimetry (see for instance [Donati \& Cameron (1997)]), the circular polarization with respect to the wavelength is recorded, which enables one to separate the components of the field, parallel or perpendicular to the line of sight. For all these applications, a sophisticated theoretical modelling is required; however, due to the huge number of electric-dipolar lines in hot plasmas, the exact calculation becomes prohibitive, and even useless due to the coalescence of the lines induced by the other physical broadenings. In the present work, we propose a statistical model for the simulation of the impact of strong magnetic fields on atomic spectra containing many overlapping lines and in which Zeeman patterns are unresolved.

\section{Statistical modelling of the intensity and calculation of the moments}

The intensity, detected with an angle of observation $\theta$ with respect to the direction of the magnetic field, is given by [Godbert-Mouret et al. (2009)]:

\begin{equation}
I_{\theta}(E)=\left(\frac{1+\cos^2(\theta)}{4}\right)\left(I_{+1}(E)+I_{-1}(E)\right)+\frac{\sin^2(\theta)}{2}I_0(E).
\end{equation}

In the following, an energy level will be denoted by $\gamma J$. The quantity $J$ stands for total angular momentum and $\gamma$, which can be interpreted as the ``signature'' of the level, represents all the quantum numbers, whose coupling leads to $J$. Each electric-dipole E1 line $\gamma J\rightarrow\gamma'J'$ splits into three components associated to the selection rules $M'$=$M+q$, where the polarization $q$ is equal to 0 for $\pi$ components and to $\pm 1$ for $\sigma_{\pm}$ components. In the present work, Hanle effect is not taken into account. For a photon energy $E$, the intensity of the $q$ component of the line $\gamma J\rightarrow\gamma'J'$ (whith energy and strength $E_{\gamma J,\gamma'J'}$ and $S_{\gamma J,\gamma'J'}$) reads:

\begin{equation}\label{defiq}
I_q(E)=\sum_{M,M'}S_{M,M',q}~\Psi(E-\eline-\mu_BB~(g_{\gamma'J'}M'-g_{\gamma J}M)),
\end{equation}

where $S_{M,M',q}=C_{M,M',q}\times S_{\gamma J,\gamma'J'}$ and $C_{M,M',q}=3\troisjm{J}{1}{J'}{-M}{-q}{M'}^2$. $\Psi$ is the line profile, which takes into account the other broadening mechanisms (Stark, Doppler, \emph{etc.}). The contribution of the magnetic field to the energy of the state $\gamma J M$ is equal to $\mu_BBg_{\gamma J}M$, where $B$ is the magnitude of the magnetic field, $\mu_B$ is the Bohr magneton, and $g_{\gamma J}$ is the Land\'e factors of level $\gamma J$. The distribution $I_q(E)$ can be characterized, in the referential of $\eline$, by the centered moments of the distribution of the energies weighted by the strengths [Mathys \& Stenflo (1987)]. The $n^{th}-$order centered moment reads:


\begin{equation}
\mathcal{M}_{n,c}^{[q]}=\sum_{M,M'}C_{M,M',q}~\left(g_{\gamma'J'}M'-g_{\gamma J}M-\mathcal{M}_1^{[q]}\right)^n .
\end{equation}

In this expression, $\mathcal{M}_1^{[q]}$ is the center-of-gravity of the strength-weighted line energies (relative to $\eline$ and in units of $\mu_BB$):

\begin{equation}\label{defge}
\mathcal{M}_1^{[q]}=\frac{q}{4}~\left(2(g_{\gamma J}+g_{\gamma'J'})\right.
\left.+(g_{\gamma J}-g_{\gamma'J'})(J-J')(J+J'+1)\right)\equiv q~g_e,
\end{equation}

and $\mathcal{M}_{2,c}^{[q]}$ is the variance (in units of $(\mu_BB)^2$). It is useful to introduce the reduced centered moments [Kendall \& Stuart (1969)] defined by:

\begin{equation}
\alpha_n^{[q]}=\left(\frac{\mathcal{M}_{n,c}^{[q]}}{\sqrt{\mathcal{M}_{2,c}^{[q]}}}\right)^n.
\end{equation}

The use of $\alpha_n^{[q]}$ instead of $\mathcal{M}_n^{[q]}$ allows one to avoid numerical problems due to the occurence of large numbers. The first values are $\alpha_0^{[q]}=1$, $\alpha_1^{[q]}=0$ and $\alpha_2^{[q]}=1$. The distribution $I_q(E)$ is therefore fully characterized by the values of $\mathcal{M}_1^{[q]}$, $\mathcal{M}_{2,c}^{[q]}$ and of the high-order moments $\alpha_n^{[q]}$ with $n>2$. The first four moments are often sufficient to capture the global shape of the distribution [Pain et al. (2009)] and $\alpha_3^{[q]}$ and $\alpha_4^{[q]}$ (see table \ref{tab1}), named {\it skewness} and {\it kurtosis}, quantify respectively the asymmetry and sharpness of the distribution ($\alpha_4^{[q]}=3$ for a Gaussian). The moments can be easily derived using Racah algebra [Landi Degl'Innocenti (1985), Mathys \& Stenflo (1987), Bauche \& Oreg (1988)]. 

\begin{table}[ht]
\begin{center}
\begin{tabular}{|c|c|c|c|} 
 & $J'=J$ & $J'=J+1$ & $J'=J-1$ \\\hline\hline
$\alpha_3~[\sigma_+]$ & 0 & $\frac{2\sqrt{5}}{3\sqrt{3}}\frac{J+1}{\sqrt{J(J+2)}}$ & $-\frac{2\sqrt{5}}{3\sqrt{3}}\frac{J}{\sqrt{J^2-1}}$ \\ 
$\alpha_4~[\sigma_+]$ & $\frac{5}{7}\left(\frac{12J(J+1)-17}{4J(J+1)-3}\right)$ & $\frac{5}{21}\left(\frac{13J(J+2)-4}{J(J+2)}\right)$ & $\frac{5}{21}\left(\frac{13J^2-17}{1-J^2}\right)$ \\
$\alpha_4~[\pi]$ & $\frac{25}{7}\left(\frac{3\{(J+2)J^2-1\}J+1}{\{1-3J(J+1)\}^2}\right)$ & $\frac{5}{7}\left(\frac{3J(J+2)-2}{J(J+2)}\right)$ & $\frac{5}{7}\left(\frac{3J^2-5}{J^2-1}\right)$ \\ 
\end{tabular}
\end{center}
\caption{Values of $\alpha_3$ and $\alpha_4$ of the $\sigma_+$ anf $\pi$ components of E1 lines ($\alpha_3$=0 for the $\pi$ component since it is symmetric).}\label{tab1}
\end{table}

\section{Taylor series expansion and diagnostic of the magnetic field}

Assuming, for simplicity, that $\Psi$ is a Gaussian with a variance $v$ (which is exact only for Doppler broadening), the quantity $I_q(E)$ of equation (\ref{defiq}) can be expressed [Mathys \& Stenflo (1987)] as a second-order Taylor series (Rodrigues' formula):
 
\begin{eqnarray}\label{iq2}
I_q(E)&=&S_{\gamma J,\gamma'J'}\Psi(E-\eline)\left\{\vphantom{\frac{\left(v-\left(E-\eline\right)^2\right)}{2v^2}}1+\mu_BB\;\mathcal{M}_1^{[q]}\frac{(E-\eline)}{\sqrt{v}}\right.\nonumber\\
& &-(\mu_BB)^2\left[\mathcal{M}_{2,c}^{[q]}+\left(\mathcal{M}_1^{[q]}\right)\right]^2\left.\frac{\left[v-\left(E-\eline\right)^2\right]}{2v^2}\right\}.
\end{eqnarray}

The Taylor-Series method is valid for $\mu_B B\lesssim\sqrt{v}$, but breaks down if $\mu_B B$ becomes much larger than $\sqrt{v}$.  However, assuming the knowledge of the variance $v$ of the other broadening mechanisms, expression (\ref{iq2}) enables one to estimate the magnitude of the magnetic field from the measurement of the full width at half maximum (FWHM) of the line:

\begin{equation}
B=\frac{1}{\mu_B}\sqrt{\frac{v(1-2e^{-\delta^2/2})}{C(\theta)\left(1-2e^{-\delta^2/2}(1-\delta^2)\right)}},
\end{equation}

where $\delta$=FWHM$/(2\sqrt{v})$, $C(\theta)=A(\theta)~\left(\mathcal{M}_1^{[\sigma_+]}\right)^2+\mathcal{M}_{2,c}^{[\sigma_+]}+B(\theta)\mathcal{M}_{2,c}^{[\pi]}$, $A(\theta)=\left(1+\cos^2(\theta)\right)/4$ and $B(\theta)=\sin^2(\theta)/4$.

\begin{figure}[ht]
\begin{center}
\vspace{1cm}
\includegraphics[width=10cm]{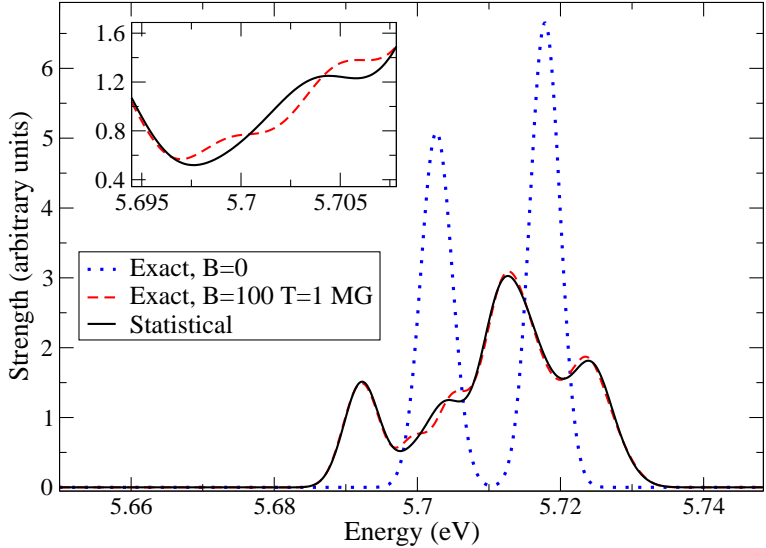}
\end{center}
\caption{Effect of a 1 MG magnetic field on triplet transition $1s2s~^3S\rightarrow 1s2p~^3P$ of C V. The Doppler full width at half maximum (FWHM) is taken to be 0.005 eV. Since, for a Gaussian, FWHM=$2\sqrt{2\ln(2)v}$, this corresponds to a variance $v\approx 4.5\;10^{-6}$ eV$^2$. The observation angle $\theta$ is such that $\cos^2(\theta)=1/3$.}
\label{fig1}
\end{figure}

\section{Fourth-order Gram-Charlier expansion series}

In case of a detailed transition array, the Zeeman broadening of a line can be represented by a fourth-order A-type Gram-Charlier expansion series: 

\begin{eqnarray}\label{newpro}
\Psi_Z(E-\eline)&=&\sum_{q=-1}^1\frac{\exp\left(-\frac{y_q^2}{2}\right)}{\mu_BB\sqrt{2\pi\mathcal{M}_{2,c}^{[q]}}}\left(1-\frac{\alpha_3^{[q]}}{2}\left(y_q-\frac{y_q^3}{3}\right)\right.\nonumber\\
& &\left.+\frac{(\alpha_4^{[q]}-3)}{24}(3-6 y_q^2+y_q^4)\right),\nonumber\\
\end{eqnarray}

where $y_q=(E-\eline-q~g_e~\mu_BB)/(\mu_BB\sqrt{\mathcal{M}_{2,c}^{[q]}})$, the coefficient $g_e$ of the line $\gamma J\rightarrow\gamma'J'$ being defined in equation (\ref{defge}). If the values of $g_{\gamma J}$ and $g_{\gamma'J'}$ are unknown, we suggest to replace $g_e$ by its average value in LS coupling $\bar{g}_e$, which can be roughly estimated, knowing the distribution of spectroscopic terms $Q(S,L)$ [Gilleron \& Pain (2009)], by:

\begin{equation}\label{varta}
\bar{g}_e=\sum_{S,L,J}\;\sum_{L',J'}\;Q(S,L')\;g_e(S,L,J,L',J')\;\Delta(L,L',J,J'),
\end{equation}

where $\Delta(L,L',J,J')$ stands for the selection rules: $|L-1|\leq L'\leq L+1$ avoiding $L'=L=0$ and $|J-1|\leq J'\leq J+1$ avoiding $J'=J=0$. The notation $g_e(S,L,J,L',J')$ means $g_e$ in which the Land\'{e} factors are estimated in LS coupling:

\begin{equation}\label{landels}
g_{SLJ}=\frac{3}{2}+\frac{S(S+1)-L(L+1)}{2J(J+1)}.
\end{equation}

Figure \ref{fig1} shows, in the case of carbon 4+ (C V in spectroscopic notation) for a 1 MG magnetic field, that Gram-Charlier series (\ref{newpro}) provides a good depiction (see the black curve named "Statistical" in the legend) of the triplet transition $1s2s~^3S\rightarrow 1s2p~^3P$. 

\section{Conclusion}

We proposed a statistical modelling of Zeeman patterns that can be easily implemented in opacity codes, together with a simple procedure for the diagnostic of magnetic fields.


\begin{thebibliography}{99}

\bibitem{ALECIAN02} Alecian G. and Stift M.J. 2002, Astron. Astrophys., {\bf 387}, 271 

\bibitem{ZEEMAN1897} Zeeman, P. 1897, Ap. J., {\bf 5}, 332

\bibitem{DONATI97} Donati J.-F. and Cameron A.C. 1997, Mon. Not. R. Astron. Soc., {\bf 291}, 1

\bibitem{GODBERT09} Godbert-Mouret L. \emph{et al.} 2009, High Energy Density Phys., {\bf 5}, 162

\bibitem{MATHYS87} Mathys G. and Stenflo J.O. 1987, Astron. Astrophys., {\bf 171}, 368

\bibitem{KENDALL69} Kendall M.G. and Stuart A., {\it Advanced Theory of Statistics} (Hafner, New York, 1969)

\bibitem{PAIN09} Pain J.-Ch. \emph{et al.} 2009, High Energy Density Phys., {\bf 5}, 294

\bibitem{LANDI85} Landi Degl'Innocenti E. 1985, Solar Phys., {\bf 99}, 1

\bibitem{BAUCHE88} Bauche J. and Oreg J. 1988, J. Physique Colloque C1, {\bf 49}, 263

\bibitem{GILLERON09} Gilleron F. and Pain J.-C. 2009, High Energy Density Phys., \textbf{5}, 320

\end{thebibliography}
\end{document}